\documentclass[12pt]{article}
\usepackage{latexsym} \usepackage{epsf}
\usepackage{epsfig}
\usepackage{a4}
\usepackage{amsfonts}
\usepackage[colorlinks]{hyperref}
\usepackage{amsmath,amssymb,bm,mathtools}
\usepackage{color}
\usepackage{graphicx}

\newcommand{\safeincludegraphics}[2][]{%
  \IfFileExists{#2}{\includegraphics[#1]{#2}}{%
    \fbox{\parbox[c][40mm][c]{70mm}{\centering #2}}}}
\newcommand{\ba}{\begin{array}}
\newcommand{\ea}{\end{array}}
\newcommand{\be}{\begin{equation}}
\newcommand{\ee}{\end{equation}}
\newcommand{\nn}{\nonumber}
\newcommand{\bea}{\begin{eqnarray}}
\newcommand{\ena}{\end{eqnarray}}
\newcommand{\beas}{\begin{eqnarray*}}
\newcommand{\enas}{\end{eqnarray*}}

\begin{document}
	
\begin{center}
 {\Large{Inhomogeneous Ising Model on 2D kagom\'{e}  Lattice: Fermionic field approach. }}
\end{center}

\begin{center}
{\bf Shahane A. Khachatryan{\footnote{e-mail:{\sl shah@mail.yerphi.am}}$^{,a}$},
		Zhidong Zhang{\footnote{e-mail:{\sl zdzhang@imr.ac.cn}}$^{,b}$  and
			 Ara G. Sedrakyan
			{\footnote{e-mail:{\sl sedrak@mail.yerphi.am}}}$^{,a,b}$}}\\
	
\vspace{12pt}
	
{\it $^a$ Alikhanyan National Sience Laboratory (Yerevan Physics Institute), Alikhanian Br. str. 2, Yerevan 36,  Armenia}\\
\vspace{6pt}
$^b$Shenyang National Laboratory for Material Science,
Institute of Metal Research, China Academy of Sciences,
72 Wenhua Road, Shenyang, 110016, P.R.China\\
\end{center}

\begin{abstract}
We investigate the two-dimensional inhomogeneous Ising model (2DIM) on the kagom'e lattice by mapping it onto a particular non-symmetric eight-vertex model and constructing the corresponding $R$-matrix. Using a fermionic representation, we evaluate the partition function and derive explicit expressions for the main thermodynamic quantities. In the thermodynamic limit, we obtain an exact equation for the critical surface determining the phase transition of the model. We also calculate the free energy, specific heat, and spontaneous magnetization in the ferromagnetic case.

Furthermore, we show that when one or two coupling constants vanish, the model reduces, respectively, to the square-lattice and one-dimensional Ising models. In both limits, our results reproduce the corresponding exact critical couplings and free energies.

\end{abstract}

\section{Introduction} 

The Ising model is one of the most fundamental models in statistical mechanics for studying phase transitions and critical phenomena. In this model, each lattice site carries a spin variable $s_i=\pm 1$, representing two possible magnetic orientations. Spins interact with their nearest neighbors, and the collective behavior of these interactions determines the macroscopic magnetic properties of the system.

In two dimensions, the Ising model has been extensively studied on various lattice geometries. The exact solution of the model on a square lattice by Lars Onsager in 1944, \cite{Onsager-1944, Kaufman-2}, established a cornerstone result in the theory of critical phenomena. However, when the lattice geometry changes, the physical properties and analytical treatment of the model may differ significantly. After Onsagers solution of the two dimensional Ising Model (2DIM) the integrability of two dimensional
statistical or quantum chain models \cite{Baxter-book, Fadeev-book} become an interesting topic in theoretical physics, which gain more and more applications
in modern low-dimensional condensed matter problems.

 The exact-solution program is also tightly connected with duality, lattice transformations, Pfaffian/dimer methods, and fermionization: the Kramers--Wannier construction fixes the square-lattice critical point, triangular and honeycomb nets provide canonical non-square benchmarks, planar Ising models can be reduced to Pfaffian or dimer problems, and the free-fermion/eight-vertex formulation gives the natural vertex-model language for the present construction \cite{KramersWannier-1941,Houtappel-1950,Wannier-1950,SchultzMattisLieb-1964,Kasteleyn-1963,Fisher-1966,FanWu-1970,Baxter-1972,KadanoffCeva-1971}.

 The transfer-matrix construction used below is aligned with Baxter's zero-field eight-vertex solution: the partition function, special transfer-matrix eigenvectors, and equivalence to generalized ice-type/Ising formulations provide the natural algebraic setting for the present kagom\'{e} $R$-matrix. Modern free-fermion developments additionally connect this point to bipartite dimers, $Z$-invariant structures, and direct free-fermion formulations of honeycomb, triangular, and kagom\'{e} Ising models \cite{Baxter-1972,Baxter-1973-I,Baxter-1973-II,Melotti-2021,LiWangYang-2025}.

One particularly interesting geometry is the kagom\'{e} lattice, a two-dimensional lattice composed of corner-sharing triangles. The kagom\'{e} lattice has a lower coordination symmetry compared with the square lattice and contains triangular units that can produce geometric frustration, especially when the spin interactions are antiferromagnetic. Because of this feature, the kagome lattice has become an important structure for studying frustrated magnetism and exotic magnetic states in condensed-matter physics.

For the ferromagnetic interaction case, the kagom\'e -lattice Ising model can be solved exactly through lattice-transformation techniques such as the star–triangle transformation, which maps the kagom\'e lattice to the honeycomb lattice. This mapping allows the determination of the system’s critical temperature and thermodynamic properties. The theoretical analysis of this system was developed in early works by K.Kano, S.Naya, I. Syozi and H.Nakano \cite{Kano-1953,Syozi-1955,Syozi-1960}, who studied decorated lattices and their transformations in the context of the Ising model. In two previous papers \cite{Matveev-1995,Kassan-Ogly-2023} the 2DIM was considered also on other
heteropolygonal lattices.

The study of the Ising model on the kagom\'{e} lattice is therefore important for several reasons. First, it provides insight into how lattice geometry affects phase transitions in two-dimensional systems. Second, it serves as a theoretical framework for understanding frustrated magnetic materials, many of which possess kagom\'{e}-type structures. Finally, the model continues to play a role in modern research on strongly correlated systems, spin liquids, and complex magnetic ordering.

 Exact mappings on closely related decorated or triangular-kagom\'{e} geometries further show that summing over internal spins can produce effective kagom\'{e} or honeycomb models with analytically tractable critical manifolds \cite{ZhengSun-2005}.

Recent kagom\'{e}-specific developments include exact analysis of interaction-generated frustration on a star kagomelike recursive lattice, quantum-simulation studies of the transverse-field kagom\'{e} Ising antiferromagnet, and finite-temperature phase diagrams of decorated kagom\'{e} Ising systems with competing interactions \cite{JurcisinovaJurcisin-2021,Narasimhan-2024,MutailamovMurtazaev-2025}. Recent review on models on kagom\'e lattice one can find in \cite{DiSante-2026}.

In this work, we present an interpretation of 2DIM on the inhomogeneous kagom\'{e} lattice \cite{Kano-1953,Syozi-1955} as a particular case of a generalized XYZ model. For the case of a regular lattice, the exact correspondence between the 2DIM and an inhomogeneous integrable XYZ model was established in \cite{KhS1,KhS2}, where the partition function of the classical model was reformulated as a trace over a product of appropriate R-matrices.

In the present work, we construct the corresponding R-matrix for the inhomogeneous 2D Ising model on the kagom\'{e} lattice (Section 2), and fermionize it using the technique developed in \cite{Sedrakyan-1998,KhS1,KhS2}.  The formulation in terms of Grassmann variables has also been applied in \cite{KhSSR,KhSSP}. This technique allows us to derive the partition function in the Fourier transform basis as a product of the determinants   (Section 3).

A central result then for the kagom\'{e} lattice model is the determination of the parametric surface of critical points, which is obtained in Section 4. For homogeneous case the critical value for coupling was known -
$\sinh[2 J_c]=(4/3)^{1/4}$, and we have reproduced this formulae.  We show here that for the inhomogeneous case the critical points belong to the two-dimensional 
critical surface, depicted in the three dimensional space of the real coupling parameters, formulated by this equation: $[\sum^3_{k=1} \cosh[2J_k]-
\prod_{k=1}^3 \sinh[2J_k]-\prod_{k=1}^3 \cosh[2J_k]]$.  We also present the exact expressions for the free energy as an integral, heat capacity in the thermodynamic limit (Section 4) and  the  expression of the spontaneous magnetization in the case of a homogeneous lattice (Section 5). The results show how
lattice geometry and coupling inhomogeneity modify the critical behavior
compared to the standard two-dimensional Ising model.

\section{Partition function: $R$-matrix}

 The kagom\'{e} lattice is demonstrated on the Fig.1. It is constructed by periodic disposed triangles and hexagons. The spins $s_\alpha=\pm 1$ of the Ising model are situated on the vertices of the lattice and only nearest neighbor interactions are considered. The spin-spin interactions for the inhomogeneous case are described by three different couplings $J_i$, $i=1,2,3$, attached to the links oriented by three different directions. The
 partition function looks like
 \bea
\label{A1}
 Z=\sum_{s_{\alpha\beta\gamma}}\prod_{\alpha\beta\gamma}e^{J_1 s_{\alpha}s_{\beta'}+J_2 s_\beta s_\gamma+J_3 s_\alpha s_\gamma}.
 \ena

\begin{figure}[ht]
\unitlength=11pt
\begin{picture}(100,15)(-3,-1)

\newsavebox{\rw}

\sbox{\rw}{\begin{picture}(3,3)
\put(0,0){\line(1,1){3}}\put(0,3){\line(1,-1){3}}
\put(0,0){\line(1,0){3}}\put(0,3){\line(1,0){3}}\put(1.5,1.5){\circle*{0.25}}
\put(0,0){\circle*{0.25}}\put(0,3){\circle*{0.25}}\put(3,0){\circle*{0.25}}\put(3,3){\circle*{0.25}}
\end{picture}}
\multiput(0,-1)(0,6){3}{\usebox{\rw}}\multiput(6,-1)(0,6){3}{\usebox{\rw}}
\multiput(12,-1)(0,6){3}{\usebox{\rw}}\multiput(18,-1)(0,6){3}{\usebox{\rw}}
\multiput(3,2)(0,6){2}{\usebox{\rw}}\multiput(9,2)(0,6){2}{\usebox{\rw}}
\multiput(15,2)(0,6){2}{\usebox{\rw}}\put(1.5,8.5){\scriptsize${J_1}$}\put(4.3,7.2){\scriptsize${J_1}$}
\put(3.1,9.2){\scriptsize${J_2}$}\put(5.4,10){\scriptsize${J_2}$}\put(5.5,8.9){\scriptsize${J_3}$}
\put(5.9,6.8){\scriptsize${J_3}$}
\put(-1.2,2.5){$\scriptstyle(2i,2j)$}\put(-2,-2){$\scriptstyle(\!2i\!+\!1,\!2j\!-\!1\!)$}
\put(3,1){$\scriptstyle(\!2i\!+\!1,2j\!+\!1\!)$}\put(3,-2){$\scriptstyle(2i+2,2j)$}
\put(25.5,5){${W}_{\alpha\beta}^{\alpha'\beta'}$=}
\put(30,4){\line(1,1){3}}\put(30,7){\line(1,-1){3}}
\put(30,4){\line(1,0){3}}\put(30,7){\line(1,0){3}}\put(31.5,5.5){\circle*{0.25}}
\put(29,4){$\beta$}\put(29,7){$\alpha$}\put(33.5,4){$\alpha'$}
\put(33.5,7){$\beta'$}
\put(30.3,5.5){\scriptsize$\gamma$}
\put(30,4){\circle{0.25}}\put(30,7){\circle{0.25}}\put(33,4){\circle{0.25}}\put(33,7){\circle{0.25}}
\end{picture} \caption{kagom\'{e} lattice}\label{fig1}
\end{figure}
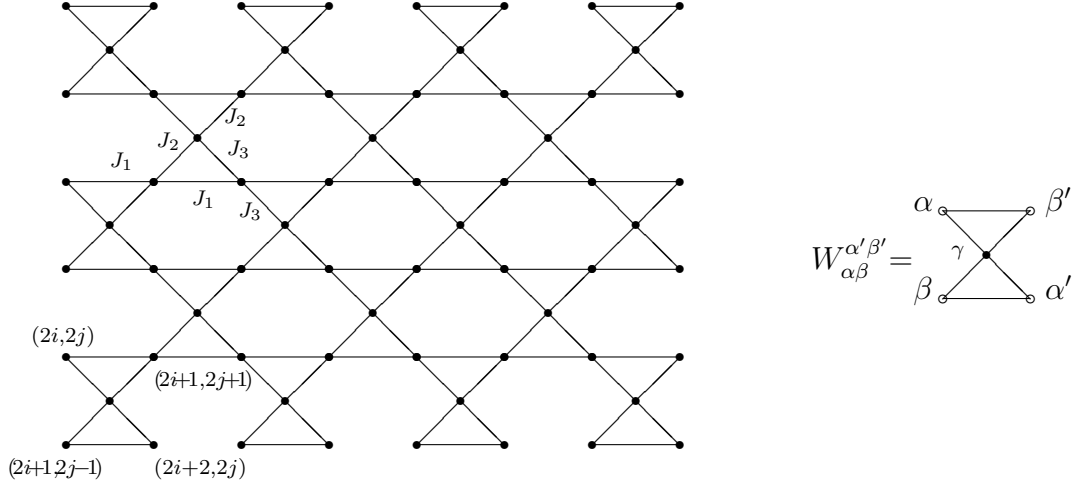

 However it is possible to represent the lattice as a checkerboard picture with square cells, constructed by means of two triangles as it is shown in the Fig. \ref{fig1} by $W_{ij}$.

   In this formulation the partition function can be rewritten as
\bea
Z=\sum_{s_{\alpha\beta}}\prod_{\alpha\beta}W_{\alpha\beta},\quad W_{\alpha\beta}=\sum_{s_\gamma=\pm 1}e^{J_1[s_{\alpha}s_{\beta'}+s_{\beta}s_{\alpha'}]+J_2 [s_\beta+s_{\beta'}] s_k+J_3 [s_\alpha+s_{\alpha'}] s_\gamma}. \label{partw}
\ena
As for the original kagom\'{e} lattice there  is rotational symmetry at the centers of the hexagons interchanging the axes, and the model has  symmetry in respect to the interchange of the parameters $J_i$, $i=1,2,3$, we could define the $R$-matrices by three different ways. In the Fig.1 we have separated the direction along the hopping $J_1$. Here we shall consider large lattice with periodic boundary conditions along the axes described by the hopping parameters $J_{2,3}$. Then, as in \cite{KhS1}, the partition function can be written as the product of the transfer matrices, defined as
\bea
Z=tr_{\{s_{(2i+1,1)}\}_{i=1,...,N}}\prod_{j}\tau_j,\quad \tau_j=tr_{s_{(0,2j)}}\prod_i R_{(2i,2j)(2i+1,2j-1)}^{(2i+2,2j)(2i+1,2j+1)}, \label{trR}
\ena
Below we shall consider the notation of the indexes as $i=0,...,N-1$, $j=1,...,N$, and the following
boundary conditions $s_{p,k+N}=s_{p,k}$, $s_{p+N,k}=s_{p,k}$.

Note that in the early work \cite{Kano-1953}, a different method was used to evaluate the partition function for the homogeneous model. The advantage of the present formulation (\ref{partw}, \ref{trR}) is that it allows one to interpret the general model as an eight-vertex model with an inhomogeneous
R-matrix, which coincides with the weight function $ W_{ij}$
up to a local unitary transformation, as shown in \cite{KhS1}.

By inserting at each vertex the identity operator $I=U U^{-1}$
 with the unitary operator $U=\frac{1}{\sqrt{2}}\left({}^{1-1}_{1\;\;1}\right)$,
the partition function can be rewritten as
\bea
Z=\sum_{s_{\alpha\beta}}\prod_{\alpha\beta}R_{\alpha\beta},\quad R_{\alpha\beta}=U^{-1}_\alpha U^{-1}_\beta W_{\alpha\beta} U_{\alpha'}U_{\beta'}. \label{partr}
\ena
One can easily calculate $W_{\alpha \beta }$  according to formula  \ref{partw} and  represent it as follows:  $\qquad  W=$
\bea
2\!\left(\!\!\ba{cccc}e^{2J_1}\cosh{2[J_2+\!J_3]}&\cosh{2J_2}&\cosh{2J_3}&e^{-2J_1}\\
\cosh{2J_2}&e^{-2J_1}\cosh{2[J_2-\!J_3]}&e^{2J_1}&\cosh{2J_3}\\\cosh{2J_3}
&e^{2J_1}&e^{-2J_1}\cosh{2[J_2-\!J_3]}&\cosh{2J_2}\\
e^{-2J_1}&\cosh{2J_3}&\cosh{2J_2}&e^{2J_1}\cosh{2[J_2+\!J_3]}\ea\!\!\right)
\ena

The corresponding $R$-matrix, maximally simplified by applying local unitary
transformations $U$ at each vertex, takes the standard form of the eight-vertex matrix.
\bea R=\left(\ba{cccc}R_{00}^{00}&0&0&R_{00}^{11}\\
0&R_{01}^{01}&R_{01}^{10}&0\\
0&R_{10}^{01}&R_{10}^{10}&0\\
R_{11}^{00}&0&0&R_{11}^{11}
\ea\right)\ena
with the following matrix elements:
\bea \label{R1}
R_{00}^{00}&=&8\left(\cosh{J_1}\cosh{J_2}\cosh{J_3}+\sinh{J_1}\sinh{J_2}\sinh{J_3}\right)^2,\\
R_{11}^{11}&=&8\left(\sinh{J_1}\cosh{J_2}\cosh{J_3}+\cosh{J_1}\sinh{J_2}\sinh{J_3}\right)^2,\\
R_{00}^{11}&=&R_{11}^{00}=2\left(\cosh{2J_2}\cosh{2J_3}-1\right)\sinh{2J_1}+2\cosh{2J_1}\sinh{2J_2}\sinh{2J_3},\\
R_{01}^{10}&=&R_{10}^{01}=2\left(\cosh{2J_2}\cosh{2J_3}+1\right)\sinh{2J_1}+2\cosh{2J_1}\sinh{2J_2}\sinh{2J_3},\\
R_{01}^{01}&=&8\left(\sinh{J_3}\cosh{J_2}\cosh{J_1}+\cosh{J_3}\sinh{J_2}\sinh{J_1}\right)^2,\\
R_{10}^{10}&=&8\left(\sinh{J_2}\cosh{J_1}\cosh{J_3}+\cosh{J_2}\sinh{J_1}\sinh{J_3}\right)^2.
\ena

The first property we have verified is the free-fermion condition, characteristic of the two-dimensional Ising model, which is found to be satisfied in this case as well.
\bea
R_{00}^{00}R_{11}^{11}-R_{00}^{11}R_{11}^{00}=R_{01}^{10}R_{10}^{01}-R_{01}^{01}R_{10}^{10}.
\ena
This give us possibility to represent in Section 3 the action of the model as a quadratic form of
Grassmann variables and write the partition function as a continual integral over them
by following the technique developed in \cite{Sedrakyan-1998,KhS1,KhS2}. After that we explore the physical characteristics of the model, namely, we find the surface
of critical points, calculate thermal capacity and magnetization.


We consider periodic boundary conditions in the same way, as in the work \cite{KhS1}.

 \section{Fermionic field representation}

 In Ref.~\cite{KhS1}, the fermionic realization of the partition function of the eight-vertex model was employed for its analysis. In the present work, we apply those results to the particular case defined by Eq.~(\ref{R1}).

 The two-dimensional spin states at each lattice site $(i,j)$ are represented in fermionic form as basis elements of a two-dimensional Fock space, $|0\rangle_{ij}$ and $|1\rangle_{ij}$. These states satisfy
 \[
 c_{ij}|0\rangle_{ij}=0, \qquad c^+_{ij}|0\rangle_{ij}=|1\rangle_{ij},
 \]
 where $c_{ij}$ and $c^+_{ij}$ are fermionic annihilation and creation operators obeying the anticommutation relation $\{c,c^+\}_+=0$, corresponding to scalar fermions.

 It should be noted that in Section II of Ref.~\cite{KhS1}, the $R$-matrix was written in the so-called ``check'' form, although this was not explicitly emphasized there; its role is clarified in Section III of that work. Applying Eqs.~(3.1)--(3.3) from \cite{KhS1}, we obtain
 \bea
 \mathcal{R}_{12}=R_{\alpha\beta}^{\alpha'\beta'}
 |\beta\rangle_2|\alpha\rangle_1{}_2\langle\beta'|{}_1\langle\alpha'|
& =&R_{\alpha\beta}^{\alpha'\beta'}
 (-1)^{p(\alpha)p(\beta')}
 |\beta\rangle_2\langle\beta'||\alpha\rangle_1\langle\alpha'|,\\
 |\alpha\rangle_k\langle\alpha'|_{(\alpha,\alpha'=0,1)}
 &=&{\left({\ba{cc}[1-c^+_k c_k]&c_k\\c^+_k&[c^+_k c_k]\ea}\right)}
 \ena
 Here, $p(\alpha)$ denotes the fermionic parity of the state, defined as $p(\alpha)=\alpha$, with $\alpha=0,1$.

 Substituting these expressions, we obtain
 \bea
 \mathcal{R}_{12}&=&R_{00}^{00}+(R_{01}^{01}-R_{00}^{00})c^+_1 c_1+(R_{10}^{10}-R_{00}^{00})c^+_2 c_2+
 R_{01}^{10}c^+_1 c_2+R_{10}^{01}c^+_2 c_1\nn\\
&+& R_{00}^{11}c^+_2 c^+_1+R_{11}^{00}c_2 c_1+
 +[R_{00}^{00}-R_{10}^{10}-R_{01}^{01}-R_{11}^{11}]c^+_2 c_2 c^+_1 c_1.
 \ena

For $R$-matrices satisfying the above-mentioned free-fermion condition, the operator $\mathcal{R}_{12}$ admits a particularly simple representation in terms of an exponential of a quadratic form in fermionic operators. More precisely, it can be written as the exponential of a local (cell) action $\mathcal{A}_{12}(c^+,c)$, which is quadratic in the fermionic creation and annihilation operators. Here, the symbol $::$ denotes normal ordering with respect to the fermionic operators.
\bea
\label{R2}
\mathcal{R}_{12}&=&
R_{00}^{00} : e^{\mathcal{A}_{12}(c^+,c)} :,
\ena
where the quadratic form $\mathcal{A}_{12}(c^+,c)$ is given by
\bea
\label{A12}
\mathcal{A}_{12}(c^+,c)&=&\left(\frac{R_{01}^{01}}{R_{00}^{00}}-1\right)c^+_1 c_1+
\left(\frac{R_{10}^{10}}{R_{00}^{00}}-1\right)c^+_2 c_2+\nn\\
&&\frac{R_{01}^{10}}{R_{00}^{00}}c^+_2 c_1+\frac{R_{10}^{01}}{R_{00}^{00}}c^+_1 c_2+
\frac{R_{00}^{11}}{R_{00}^{00}}c^+_2 c^+_1+\frac{R_{11}^{00}}{R_{00}^{00}}c_2 c_1.
\ena
This representation makes explicit the quadratic (free-fermion) structure of the model and provides a convenient starting point for further analytical treatment, in particular for constructing the corresponding Grassmann path integral.

The indexing of the states on the lattice can be introduced as follows (see Fig.~1):

 %
 %
 \begin{figure}[ht]\unitlength=8pt
 \begin{picture}(100,20)(-15,0)	\safeincludegraphics[width=70mm]{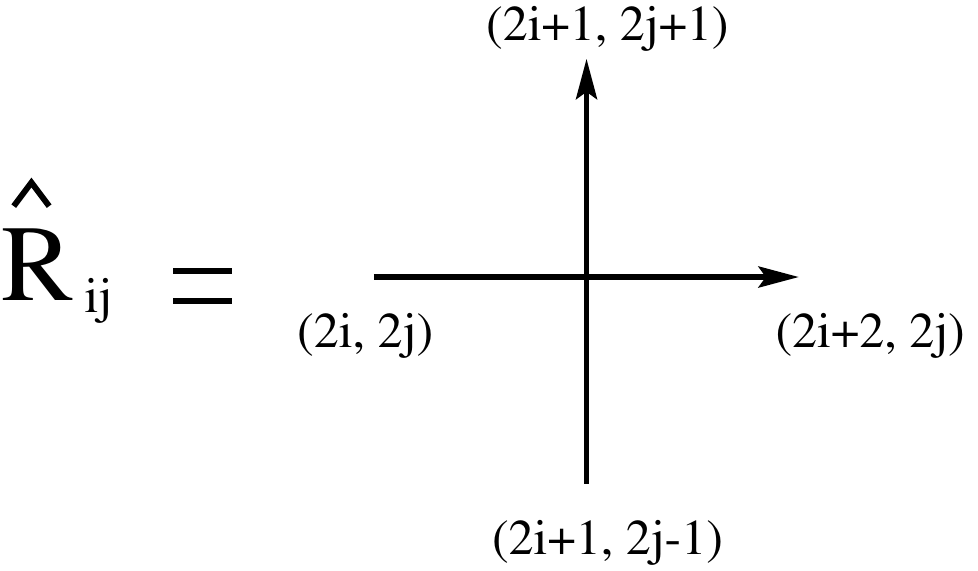}
 \end{picture}
 	\caption{R-operator}
 	\label{fig2}
 \end{figure}

The periodic boundary conditions imposed on the spin variables of the two-dimensional lattice translate into antiperiodic boundary conditions for the corresponding fermionic variables. This must be taken into account when passing to Grassmann field variables $\psi,\bar{\psi}$ in order to represent the trace in the partition function as a functional integral.

The fermionic variables $\psi,\bar{\psi}$ are associated with coherent states (eigenstates) of the fermionic annihilation operators,
\[
|\psi\rangle=e^{c^+ \psi}|0\rangle, \qquad c|\psi\rangle=\psi|\psi\rangle.
\]
To avoid repeating the detailed derivations presented in Ref.~\cite{KhS1}, we provide here only the key formulas, using a compact notation.

\bea
Z&=&tr \prod_{ij} R_{ij}=[R_{00}^{00}]^{N\times N}\int D\bar{\psi}D\psi e^{-\sum_{ij}{A}(\bar{\chi},\psi)_{ij}},\\
\psi_{(i,j)}&=&\left(\ba{c}\psi(2i,2j)\\\psi(2i+1,2j-1)\ea\right),\quad
\bar{\chi}_{(i,j)}=\left(\ba{cc}\bar{\psi}(2i+2,2j),&\bar{\psi}(2i+1,2j+1)\ea\right)\\
\psi_{(i,j)}&=&-\psi_{(i+N,j)},\quad\psi_{(i,j)}=-\psi_{(i,j+N)},\quad \bar{\chi}_{(i,j)}
=-\bar{\chi}_{(i+N,j)},\quad\bar{\chi}_{(i,j)}=-\bar{\chi}_{(i,j+N)} \qquad
\ena
The above relations encode the antiperiodic boundary conditions for the fermionic fields on the torus.

The fermionic action can be written as
\bea
\label{FA}
&-&A(\bar{\psi},\psi)_{ij}
=-\sum_{ij}\left[\bar{\psi}(2i,2j){\psi}(2i,2j)+\bar{\psi}(2i+1,2j+1){\psi}(2i+1,2j+1)\right]\nn\\\nn
&+&\sum_{j}\bar{\psi}(2N,2j){\psi}(0,2j)+\sum_{i}\bar{\psi}(2N+1,2j+1){\psi}(1,2j+1)
+\sum_{i,j}\bar{\chi}_{(i,j)}
\left(\ba{cc}\frac{R_{01}^{01}}{R_{00}^{00}}&\frac{R_{01}^{10}}{R_{00}^{00}}\\
\frac{R_{10}^{01}}{R_{00}^{00}}&\frac{R_{10}^{10}}{R_{00}^{00}}\ea\right)\psi_{(i,j)}\\
&+&\sum_{i,j}\left[
\frac{R_{11}^{00}}{R_{00}^{00}}\bar{\psi}(2i+2,2j)\bar{\psi}(2i+1,2j+1)+
\frac{R_{00}^{11}}{R_{00}^{00}}{\psi}(2i,2j){\psi}(2i+1,2j-1)\right]
\ena
Since the action is quadratic in the Grassmann variables, the partition function can be evaluated exactly by transforming to momentum space. In this representation, the model describes free scalar fermions with periodic hopping amplitudes on a toroidal lattice.

Due to the antiperiodic boundary conditions, the Fourier expansion must be performed over half-integer (odd) momenta \cite{KhS1}:
\bea
\psi_{(i,j)}=\frac{1}{N}
\sum_{i,j}^{N}\left(
\ba{c}e^{-\frac{i\pi}{2N}\left((2n_i+1)(2i)+(2n_j+1)(2j)\right)}
\psi_1{(\frac{\pi(2n_i+1)}{2N},\frac{\pi(2n_j+1)}{2N})}\\
e^{-\frac{i\pi}{2N}\left((2n_i+1)(2i+1)+(2n_j+1)(2j-1)\right)}
\psi_2{(\frac{\pi(2n_i+1)}{2N},\frac{\pi(2n_j+1)}{2N})}\ea \right)
\ena
where $n_i,\;n_j=1,\dots,N$.

To diagonalize the action, it is convenient to redefine the fermionic Fourier modes $\psi_{k(n_i,n_j)}$, $k=1,2$, by restricting the momentum space to half of the Brillouin zone, $n_i=1,\dots,[N]/2$, $n_j=1,\dots,N$, and introducing the following relations:
\bea
\psi_{1(N-n_i,N-n_j)}\equiv-\bar{\psi}_{3(n_i,n_j)},\quad \psi_{2(N-n_i,N-n_j)}\equiv-\bar{\psi}_{4(n_i,n_j)}\nn\\
\bar{\psi}_{1(N-n_i,N-n_j)}\equiv{\psi}_{3(n_i,n_j)},\quad
\bar{\psi}_{2(N-n_i,N-n_j)}\equiv{\psi}_{4(n_i,n_j)}.
\ena
This transformation leads to the following expression for the effective fermionic action in momentum space:
%
%
\bea
\sum_{i,j}^{N,N}A(\bar{\psi},\psi)_{ij}=\sum_{n_i,n_j}^{\frac{N}{2},N}\sum_{k,r}^{4}
\bar{\psi}_{k(n_i,n_j)}\mathcal{A}_{kr(n_i,n_j)}{\psi}_{r(n_i,n_j)},
\ena
where the matrix $\mathcal{A}_{(n_i,n_j)}$ is given by
\bea\label{A-f}
\mathcal{A}_{(n_i,n_j)}\!=\!\!\left(\!\!\ba{cccc}
\frac{R_{01}^{01}}{R_{00}^{00}}e^{\frac{i\pi(2n_i+1)}{N}}-1&
\frac{R_{01}^{10}}{R_{00}^{00}}e^{\frac{i\pi(n_i+n_j+1)}{N}}&0&-
\frac{R_{11}^{00}}{R_{00}^{00}}e^{\frac{i\pi(n_i-n_j)}{N}}\\
\frac{R_{10}^{01}}{R_{00}^{00}}e^{\frac{i\pi(n_i+n_j+1)}{N}}&
\frac{R_{10}^{10}}{R_{00}^{00}}e^{\frac{i\pi(2n_j+1)}{N}}-1&
\frac{R_{11}^{00}}{R_{00}^{00}}e^{\frac{i\pi(n_j-n_i1)}{N}}&0\\
0&\frac{R_{00}^{11}}{R_{00}^{00}}e^{\frac{i\pi(n_j-n_i)}{N}}&
\frac{R_{01}^{01}}{R_{00}^{00}}e^{\frac{-i\pi(2n_i+1)}{N}}-1&
\frac{R_{10}^{01}}{R_{00}^{00}}e^{-\frac{i\pi(n_i+n_j+1)}{N}}\\
-\frac{R_{00}^{11}}{R_{00}^{00}}e^{\frac{i\pi(n_i-n_j)}{N}}&0&
\frac{R_{01}^{10}}{R_{00}^{00}}e^{-\frac{i\pi(n_i+n_j+1)}{N}}&
\frac{R_{10}^{10}}{R_{00}^{00}}e^{\frac{-i\pi(2n_j+1)}{N}}-1\ea\!\!\right).
\ena

--------------------------------------------------------------------------------------

The action in momentum space decomposes into independent blocks labeled by $(n_i,n_j)$.
For each momentum sector, it takes the form of a $4\times 4$ matrix, similar to the structure
of the eight-vertex (or IM) model on the square lattice \cite{KhS1}. As a result, the partition
function factorizes into a product of determinants:
\bea
\label{Z}
Z=[R_{00}^{00}]^{N\times N}\prod_{n_i,n_j}^{N/2,N}\mathrm{Det}{\mathcal{A}_{(n_i,n_j)}},
\ena
Each determinant can be calculated explicitly and has the simple form
\bea
\label{det}
\left[\mathrm{Det}{\mathcal{A}_{(n_i,n_j)}}\right][R_{00}^{00}]^2&=&\mathbb{A}_1
+\mathbb{A}_2\cos{[\frac{\pi(2n_i+1)}{N}]}
+\mathbb{A}_3\cos{[\frac{\pi(2n_j+1)}{N}]}\\
&+&\mathbb{A}_4\cos{[\frac{\pi(n_i+n_j+1)}{N}]}
+\mathbb{A}_5\cos{[\frac{\pi(n_i-n_j)}{N}]},\nn
\ena
where
\bea
\label{coefficients}
\hspace{-1.2cm}\mathbb{A}_1&=&4((6+3\sum_k^3\cosh{[4J_k]})+
\prod_k^3\sinh{[4J_k]}+\prod_k^3\cosh{[4J_k]}),\\ \label{coefficients-2}
\hspace{-1.2cm}\mathbb{A}_2&=&-32\sinh{[2J_2]}(\cosh{[2J_2]}\sinh{[2J_1]}\sinh{[2J_3]}+
\sinh{[2J_2]}\cosh{[2J_1]}\cosh{[2J_3]}),\\
\hspace{-1.2cm}\mathbb{A}_3&=&-32\sinh{[2J_3]}(\cosh{[2J_3]}\sinh{[2J_2]}\sinh{[2J_1]}+
\sinh{[2J_3]}\cosh{[2J_2]}\cosh{[2J_1]}),\\
\hspace{-1.2cm}\mathbb{A}_4&=&-32\sinh{[2J_1]}(\cosh{[2J_1]}\sinh{[2J_2]}\sinh{[2J_3]}+
\sinh{[2J_1]}\cosh{[2J_2]}\cosh{[2J_3]}),\\
\hspace{-1.2cm}\mathbb{A}_5&=&[R_{01}^{01}][R_{10}^{10}]-[R_{00}^{11}][R_{11}^{00}]=0.
\ena

\section{Critical points}
The partition function can be written in a compact product form by introducing the notation
$\mathbb{A}[J_1,J_2,J_3]\equiv\mathbb{A}_2$ (see Eq.~\Ref{coefficients-2}) and
$p_{ij}\equiv[\frac{\pi(n_i+n_j+1)}{N}]$:
\bea
\label{PF}
Z(J_1,J_2,J_3)
= \prod_{n_i,n_j}^{\frac{N}{2},N}\Big\{\mathbb{A}_1
+\mathbb{A}[J_1,J_2,J_3]\cos{p_{ii}}
+\mathbb{A}[J_2,J_3,J_1]\cos{p_{jj}}
+\mathbb{A}[J_3,J_1,J_2]\cos{p_{ij}}\Big\}\nn\\
\ena
The critical behavior of the system is determined by the zeroes of the determinant.
These zeroes correspond to singularities of the free energy and therefore define the
critical surface.

As in the ordinary Ising model on the square lattice, the thermodynamic limit $N\to \infty$
is controlled by the long-wavelength modes. In particular, one must consider the sector
with $n_i=0$ and $n_j=0,N-1$. In these points the zeroes of the  partition function are determined by the equation
\bea
\label{cp}
\left(\sum_{k=1}^3\cosh{[2J_k]}-
\prod_{k=1}^3\sinh{[2J_k]}-\prod_{k=1}^3\cosh{[2J_k]}\right)=0
\ena
which defines the critical surface (Fig.~3).
 All calculations presented so far are valid for arbitrary signs of the coupling constants. Therefore, in Fig.~3 we show both the positive and negative regions of the coupling-constant parameter space.
Equation (\ref{cp}) has two analytic solutions expressing $J_3$ via $J_{1,2}$:
\bea
\label{sol-crit}
J_3^{\pm}=\frac{1}{2}\log\Big[\frac{\big(\sqrt{\cosh2[J_1]}\pm \sqrt{\cosh[2 J_2]}\big)^2}{[\cosh{[2(J_1+J_2])}-1]}\Big] 
\ena 

The choice $J_{1,2,3}>0$ corresponds to the ferromagnetic model, for which we present below the isotropic limit as a particular case. Equation~(\ref{cp}) is invariant under the simultaneous sign reversal of any two coupling constants, $J_k$ and $J_p$. In contrast, if only one coupling constant or all three coupling constants change sign, the equation is no longer invariant and yields different solutions. This gives rise to the exotic frustration effects characteristic of the antiferromagnetic phases.

\begin{figure}[ht]\unitlength=8pt
	\begin{picture}(100,20)(0,0)	\safeincludegraphics[width=80mm]{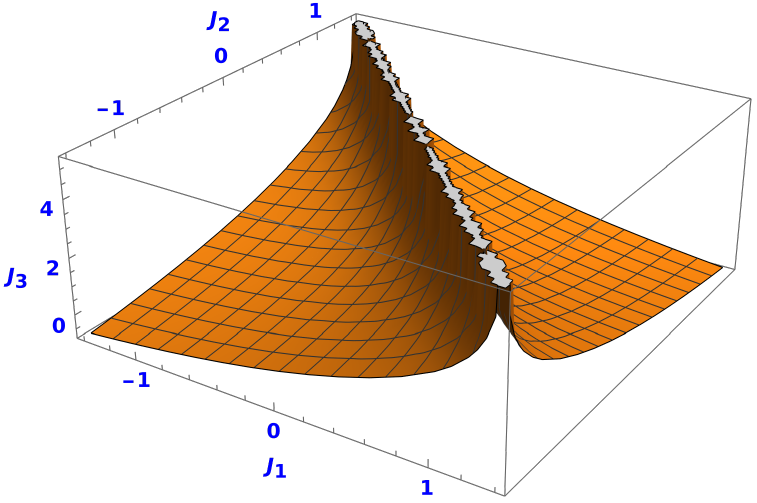}
		\safeincludegraphics[width=80mm]{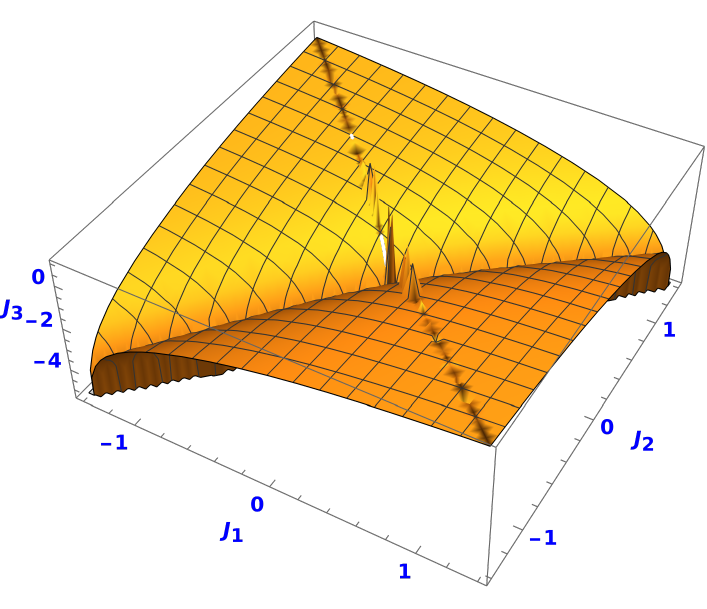}
	\end{picture}
	\caption{Critical subsurfaces on the real coordinate space \{$J_1,J_2,J_3$\}}
	\label{Surf1}
\end{figure}

The factorization of the partition function into momentum-dependent determinants is the consequence of that
the model has an effective quadratic (free-fermion–like) structure. Each momentum mode
contributes independently to the thermodynamics. The critical surface is determined by the condition that the momentum-space fermionic kernel develops a zero eigenvalue at some momentum in the Brillouin zone. Equivalently, $detA(p_0,q_0)=0$ for a certain $(p_0,q_0)$. In ferromagnetic case usually $p_0=q_0=0$, while in frustrated
	case gap may disappear in the corner of Brillouin zone.  
	This does not imply that the free energy vanishes; rather, the logarithmic integrand becomes singular, producing a nonanalyticity of the free energy in the thermodynamic limit. Since the vanishing eigenvalue corresponds to a zero fermionic mass, the fermionic spectrum is gapless at every physical point of the critical surface.
This corresponds to the closing of the excitation
gap and leads to a second-order phase transition. Compared to the square-lattice Ising model,
the form of the critical condition is modified by the kagom\'{e} lattice geometry and by the
presence of three coupling constants, but the physical mechanism remains the same.

For the isotropic case we have
\bea
\label{CP}
\left(3+\cosh{[4J_c]-2\sinh{[4J_c]}}\right)=0,\quad \sinh{2J_c}=\frac{\sqrt{2}}{\sqrt[4]{3}},
\ena
which agrees with the results of Refs.~\cite{Syozi-1955, Naya-1954,Matveev-1995} and gives the known numerical value
\bea
1/{J_c} \approx 2.14332
\ena

For other momentum values, the partition function may have zeroes at complex temperatures. These are Fisher zeroes and do not correspond to physical phase transitions on the real temperature axis.

In the limit where one of the coupling constants $J_k$ vanishes, the model reduces to the ordinary square-lattice Ising model. In the case $J[1]=0$, the equation (\ref{cp}) for critical couplings become
\bea
\label{j3}
(1 - 2 \sinh{[J_2]}^2 \sinh{[J_3]}^2)=0,
\ena
which reproduces the standard result.
 Indeed, 	from Fig.~1 it is easy to see that, upon setting $J_1=0$, the hopping terms in the action (\ref{A1}) along direction $1$ disappear, leaving a modified square (regular) lattice with additional sites of degree $2$ inserted in the middle of the links carrying couplings $J_2$ and $J_3$. The spins on these degree-$2$ sites can be summed over exactly in the partition function, yielding renormalized couplings for the ordinary Ising model on the square lattice. Namely, for each link along the $a=2,3$ directions, the two hopping terms contribute
\bea
\label{Hop1}
&&\sum_{s_0=\pm1} e^{J_a s_1 s_0 + J_a s_0 s_2}=\sum_{s_0=\pm1} \left(\cosh[J_a]+\sinh[J_a] s_1 s_0\right)
\left(\cosh[J_a]+\sinh[J_a] s_0 s_2\right)\nn\\
&=&2 \left(\cosh[J_a]^2 + \sinh[J_a]^2 s_1 s_2\right) = 2 e^{\bar J_a} \left( \cosh[\bar J_a] + \sinh[\bar J_a] s_1 s_2 \right),
\ena 
 is precisely the Boltzmann weight of the Ising model on the regular square lattice with the renormalized coupling $\bar J_a$. Comparing both sides, one immediately obtains
  	\bea
  	\label{barJ}
  	e^{2\bar J_a}=\cosh(2J_a).
  	\ena
  This equation, or equivalently the equation $\tanh{\bar J_a}=(\tanh{J_a})^2$ shows, that the resulting
	two dimensional square Ising model must be ferromagnetic. The critical equation for the square  IM with couplings $\bar{J}_{a=2,3}$ can be presented by
	$(1-\sinh{2\bar{J}_2}\sinh{2\bar{J}_3}=0)$ or, equivalently, for the ferromagnetic IM, by relations  $e^{-2\bar{J}_{2,3}}=\tanh{\bar{J}_{3,2}}$, as
	\bea\nn
	1-\sinh{[2\bar{J}]}\sinh{[2\bar{J}']}
	=\left(1-e^{2\bar{J}}\tanh[\bar{J}']\right)\left(1+e^{-2\bar{J}}\tanh[\bar{J}]\right)\cosh[\bar{J}]\cosh[\bar{J}']e^{\bar{J}-\bar{J}'}
	\ena
	Denote, duality transformation of the square
	lattice equivalent to interchanging of couplings $J_{2,3} \rightarrow J_{3,2}  $. Applying the transformation
	expression (\ref{barJ}) we find out 
\bea
(1-e^{2\bar{J}_{2}}\tanh{[\bar{J}_{3}]})=
1-\cosh{[2 J_2]}\tanh{[J_3]}^2=
\frac{1-2\sinh{[J_2]}^2\sinh{[J_3]}^2}{\cosh[J_3]^2}=0,
\ena
 coinciding with the  equation (\ref{j3}).  

It is straightforward to check also, that in the homogeneous case $J_2=J_3=J=\operatorname{arcsinh}(2^{-1/4})$, the solution of Eq.~(\ref{j3}) reproduces the well-known critical coupling of the square-lattice Ising model,
  	\bea
  	2\bar J_a
  	=\ln\left[\cosh[2\;\operatorname{arcsinh}[2^{-1/4}]]\right]
  	=\operatorname{arcsinh}[1].
  	\ena
  	Hence, the criticality condition (\ref{cp}) for the kagom'e Ising model correctly reduces to the exact critical coupling of the two-dimensional square-lattice Ising model \cite{Baxter-1972}.
  	
 We can consider also second coupling to be zero $J_2=0$, then we have one dimensional Ising limit. Taking limit $J_2 \rightarrow 0 $  in the solution of equation (\ref{j3}) we get
\bea
\label{1d}
\lim_{J_2 \rightarrow 0}  \sinh[J_3] =\lim_{J_2 \rightarrow 0} \frac{1}{2 \sinh[J_2]}=\infty,
\ena 
indicating that phase transition happened at $J_3=\infty$, equivalent to $T=0$, which is precisely
critical point of 1dIM \cite{Baxter-1972}.

\paragraph{\large Free energy per site and thermal capacity.}
We now calculate the free energy of the 2D Ising model on the kagom\'{e} lattice in ferromagnetic case. In the thermodynamic limit $N\to \infty$, the sums over momenta can be replaced by integrals. At the same time, we restore the temperature dependence by the substitution $J\to J/T$:
\bea
\frac{1}{N}\sum_{n_i}^{N/2}\to \frac{1}{\pi}\int_{0}^{\pi/2}dp,\quad
\frac{1}{N}\sum_{n_j}^{N}\to \frac{1}{\pi}\int_{0}^{\pi}dq,\qquad
J\to \frac{J}{T}.
\ena
This leads to the following integral expression for the free energy in the continuum limit:
\bea \label{F-a}
F&=&-\frac{(T \mathrm{ln} Z)}{N^2}=-\frac{T}{N^2}\sum_{(n_i,n_j)}^{N/2,N}\mathrm{ln}
\left[\mathrm{Det}{\mathcal{A}_{(n_i,n_j)}}[R_{00}^{00}]^2\right]\\\nn
&=&-\frac{T}{\pi^2}\int_0^{\pi/2}dp\int_0^{\pi}dq\ln\left[\mathbb{A}_1
	+\mathbb{A}_2\cos{[2p]}
	+\mathbb{A}_3\cos{[2q]}
	+\mathbb{A}_4\cos{[2(p+q)]}\right],
\ena
This expression has the same structure as in other exactly solvable lattice models.

In the isotropic case, the coefficients $\mathbb{A}_k$ with $k=2,3,4$ become equal. The free energy then simplifies to
\bea
\label{Fs}
\hspace{-2cm} F&=&-\frac{T}{\pi^2}\int_0^{\frac{\pi}{2}}dp\int_0^{\pi}dq \ln \left[\mathbb{A}_1
+\mathbb{A}\Big[\frac{J}{T},\frac{J}{T},\frac{J}{T}\Big]\big(\cos{[2p]}
+\cos{[2q]} +\cos{[2(p+q)]}\big)\right]\\
&=&-\frac{T}{\pi^2}\int_0^{\frac{\pi}{2}}
dp\int_0^{\pi}dq\times \ln \left[24+21 e^{-4J}+18e^{4J}+e^{12J}-32\; e^{2J}\sinh[2J]^2\cosh[2J]\right.\nn\\
&\times& \left. (\cos[2p] + \cos[2q] +\cos[2(p+q)])\right]\nn
\ena
After obtaining the free energy, the thermal capacity can be calculated in a standard way:
\bea
C=-T\frac{\partial^2 F}{\partial T^2},
\ena
This leads to expressions involving complete elliptic integrals $E$ and $K$, similar to the square-lattice Ising model, where the same method reproduces Onsager's result \cite{KhS1}. To avoid very long expressions, we consider only the isotropic case.

Introducing the notation
\bea
z\equiv e^{\frac{4J}{T}},\quad c[p,q]\equiv \cos{2q}+\cos{2p}+\cos{2(p+q)},
\ena
the heat capacity can be written as
\bea
C=\left(\frac{2 J}{ T\sinh{[\frac{4J}{T}]}}\right)^2
\left((3+\cosh{[\frac{4J}{T}]})-\frac{1}{\pi^2}\int_0^{\pi/2}dp\int_0^{\pi} dq\times\right.\nn\\
\left. \left[\frac{{45 + 291 z+ 501 z^2 + 523 z^3 + 159 z^4 + 17 z^5 - z^6 + z^7}}{z(21 + 24 z + 18 z^2 + z^4 -
	4 c[p,q] (1 - z)^2 (1 + z)^2)}\right]+\right.\\\nn
\left.\frac{1}{\pi^2}\int_0^{\pi/2}dp\int_0^{\pi} dq\left[\frac{(3+z)(3+6 z-z^2)(5+2 z+z^2)}{(21 + 24 z + 18 z^2 + z^4 -
	4 c[p,q] (1 - z)^2 (1 + z))} \right]^2\right)
\ena

After performing the integrations, we obtain the final result:
\bea
C=\frac{4J^2}{T^2\sinh{[\frac{4J}{T}]}^2}
\left(3+\cosh{[\frac{4J}{T}]}\right)-
\frac{4 J^2}{ T^2\sinh{[\frac{4J}{T}]}^2\pi }\times\quad \nn\\
\frac{18+ 93 z+ 84 z^2 + 58 z^3 + 2 z^4 + z^5}{2(5+2 z+z^2)}Elliptik{\mathbf{K}}[\frac{16(1+z)(z-1)^2}{(5+2z+z^2)^2}]\\
-\frac{4 J^2}{ T^2\sinh{[\frac{4J}{T}]}^2\pi}\frac{(3+z)^2}{2}Elliptik{\mathbf{E}}[\frac{16(1+z)(z-1)^2}{(5+2z+z^2)^2}]
\quad \nn
\ena

\paragraph{\large Reduction to the square-lattice Ising model at $J_1=0$.}

As we have mentioned earlier at $J_1=0$  (or any other coefficient) we obtain square lattice.
Lets check this statement on free energy (\ref{F-a}). At $J_1=0$ the expressions (\ref{coefficients}) for coefficients become
\bea
\label{AI}
A_1&=&16(\cosh[2 J_2]^2+1)(\cosh[2 J_3]^2+1)\nn\\
A_2&=& -32 \cosh[2 J_3] \sinh[2 J_2]^2\\
A_3&=& -32 \cosh[2 J_2] \sinh[2 J_3]^2 ,\;\;
A_4=A_5=0 \nn
\ena
and from (\ref{F-a}) we obtain ({below we skip the notation of the temperature $T$ in the expressions of $J_k/T$)	
\bea
\label{FF1}
 \hspace{-0. cm} F(0,J_2,J_3)&=& -\frac{T}{\pi^2}\int_0^{\pi/2} dp\int_0^\pi dq \ln\Big\{16\Big[ (\cosh[2 J_2]^2+1)(\cosh[2 J_3]^2+1)\nn\\
 &-& 2 \cosh[2 J_2] \sinh[2 J_3]^2 \cos[2 p]-2 \cosh[2 J_3] \sinh[2 J_2]^2 \cos[2 q] \Big] \Big \} 
\ena
The integral over p can be taken precisely giving one dimensional integral form for free energy
\bea 
\label{FF2}
F(0,J_2,J_3)&=& -\frac{T}{2 \pi}\int_0^{\pi} dq \ln\Big[8\big(f_0(q)+\sqrt{M(q)^2-4\cosh[2 J_2]\sinh[J_3]^4} \big)\Big],
\ena 
where 
\bea
\label{MM}
f_0(q)=(\cosh[2 J_2]^2+1)(\cosh[2 J_3]^2+1)-2 \cosh[2 J_3]\sinh[2 J_2]^2 \cos[2 q].
\ena
This is the usual one-dimensional integral representation of the anisotropic square-lattice Ising free energy.  Now, by using formulas (\ref{barJ}) for renormalized couplings in
reduced Ising model from kagom\'e to square lattice one will get
\bea
\label{reduced}
F(0,J_2,J_3)=F_{Ising}(\bar J_2, \bar J_3)-T \ln\Big[ 4 \sqrt{\cosh[2 J_2]\cosh[2 J_3]}\Big]\nn\\
Z(0,J_2,J_3)=\Big[ 4 \sqrt{\cosh[2 J_2]\cosh[2 J_3]}\Big]^{N^2}Z_{Ising}(\bar J_2,\bar J_3),
\ena  
which is sound with formula (\ref{barJ}).

Thus, the free-energy expression obtained from the fermionic formulation is fully consistent with the reduction formula (\ref{barJ}) and confirms that, at $J_1=0$, the model is equivalent to the anisotropic square-lattice Ising model with renormalized couplings $\bar J_2$ and $\bar J_3$.

\paragraph{\large Reduction to the Ising model on a line at $J_1=J_2=0$.}

In this case we have only two nonzero coefficients in the general free energy expression
(\ref{F-a}). Those are
\bea
\label{AI-2}
A_1&=&32(\cosh[2 J_3]^2+1)\\
A_2&=& -32 \sinh[2 J_3]^2,\;\qquad
A_3=A_4=A_5=0. \nn
\ena
Therefore, the free energy reduces to
\bea
\label{FF4}
F(0,0,J_3)&=& -\frac{T}{\pi^2}\int_0^{\pi/2} dp\int_0^\pi dq \ln\Big\{32\Big[\cosh[2 J_3]^2+1
- \sinh[2 J_3]^2 \cos[2 p] \Big] \Big \}
\ena
Since the integrand is independent of q, the q-integration gives a factor $\pi$ and
after integration over p we obtain
\bea
\label{FF5}
F(0,0,J_3)=-T \ln\Big[4(\cosh[2 J_3]+1)\Big].
\ena
After making renormalization by (\ref{barJ}) we receive
\bea
\label{FF6}
F(0,0,J_3)&=&F_{1DIM}(\bar J_3)-T \ln\Big[4 e^{\bar J_3}\Big],\nn\\
F_{1DIM}(\bar J_3)&=&-T\ln\Big[2 \cosh[\bar J_3]\Big],
\ena
where $F_{1DIM}(\bar J_3) $ is correct free energy of one dimensional Ising model.

\section{Spontaneous magnetization}

In Ref.~\cite{KhS1} we developed a method to calculate the magnetization in ferromagnetic case of two-dimensional spin models with the free-fermion property. The method is based
on a fermionic (Grassmann) representation of the model in terms of scalar
fermions and continuum functional integrals.

The spontaneous magnetization $\langle s_{(i,j)}\rangle$ can be obtained from the
long-distance behavior of the spin-spin correlation functions on an infinite
$N\times N$ lattice:
\bea
\langle\sigma_{\alpha}(i,j)\rangle^2=
\textrm{lim}_{N\to\infty}\left(\textrm{lim}_{K\to\infty}
\langle s_{(0,0)}s_{(K,K)}\rangle\right)
=\textrm{lim}_{N\to\infty}\left(\textrm{lim}_{K\to\infty}
\langle s_{(0,0)}s_{(0,K)}\rangle\right).
\ena
Thus, the spontaneous magnetization is determined by the asymptotic value of the
two-point correlation function at large separation.

In the general case, the finite-distance spin-spin correlation function in the
fermionic representation can be written as a Pfaffian. Without loss of generality,
one can fix a direction and consider even-even or odd-odd lattice sites.
For example, for even-even sites, the analysis of Ref.~\cite{KhS1} gives
\bea
\label{Gk}
G(k)=\!\langle s_{(0,2k)}s_{(0,0)}\rangle\!=\!\langle [c^+_{(0,2k)}\!+\!c_{(0,2k)}]\!
\left(\!\prod^k_{r=0}\![c_{(0,2r)}\!-\!c^+_{(0,2r)}]\![c^+_{(0,2r)}\!+\!c_{(0,2r)}]\!\right)
[c_{(0,0)}\!-\!c^+_{(0,0)}]\rangle.
\label{Gi}
\ena
Using translational invariance, the coordinates $(0,0)$ and $(0,2k)$ can be
shifted to arbitrary points $(2i,2j)$ and $(2i,2j+2k)$.

In the operator formulation, the expectation value of a spin-dependent function
$f[s_{i,j}]$ is defined as
\bea
\langle f[s_{i,j}] \rangle=Tr_{s}\left[ f[s_{i,j}] \prod_{i,j} R_{i,j}\right].
\ena
To rewrite this expression as a Grassmann integral, the fermionic operators must
be brought to normal-ordered form. The spin variable can be expressed through
fermions using a two-dimensional analog of the Jordan--Wigner transformation
(see the Appendix of Ref.~\cite{KhS1}):
\bea
s_{(i,j)}=
[c_{(i,j)}^++c_{(i,j)}]\prod_{(k,r)=(0,0)}^{(i,j)}[c^+_{(k,r)}+c_{(k,r)}][c^+_{(k,r)}-c_{(k,r)}].
\ena
The path connecting $(0,0)$ and $(i,j)$ can be chosen arbitrarily, as discussed
in Ref.~\cite{KhS1}.

Applying Wick's theorem, the correlation function (\ref{Gi}) can be written as a
Gaussian Grassmann integral:
\bea\label{gk}
G(k)=\int D\chi \, e^{\frac{1}{2}\sum_{i,j}^{2k}\mathcal{G}_{ij}\chi_i \chi_j-\sum_i^{k}\chi_{2i+1}\chi_{2i}},
\ena
where $\mathcal{G}_{ij}$ is an antisymmetric matrix. Introducing the notation
$c_{(0,2i)}\equiv \mathbf{c}_i$, its elements are given by
\bea
&\{\mathcal{G}_{(2i,2j)},\mathcal{G}_{(2i+1,2j+1)},\mathcal{G}_{(2i,2j+1)}\}=&\label{gkr}\\\nn
&\{\langle[\mathbf{c}^+_{i}+\mathbf{c}_{i}][\mathbf{c}^+_{j}+\mathbf{c}_{j}]\rangle,
\langle[\mathbf{c}^+_{i}-\mathbf{c}_{i}][\mathbf{c}^+_{j}-\mathbf{c}_{j}]\rangle,
\langle[\mathbf{c}^+_{i}+\mathbf{c}_{i}][\mathbf{c}^+_{j}-\mathbf{c}_{j}]\rangle.\}&
\ena
These matrix elements admit the following integral representation:
\bea
\mathcal{G}_{(i,j)}=\frac{(R_{00}^{00})^{N^2}}{Z}\int D\bar{\psi}D\psi \,
e^{A(\bar{\psi},\psi)}\left(x'_1\psi_{(0,2i)}+x'_2\psi_{(1,2i-1)}+x'_3
\bar{\psi}_{(2,2i)}+x'_4\bar{\psi}_{(1,2i+1)}\right)\nn \\
\times\left(x''_1\psi_{(0,2j)}+x''_2\psi_{(1,2j-1)}+x''_3
\bar{\psi}_{(2,2j)}+x''_4\bar{\psi}_{(1,2j+1)}
\right).
\label{gij}
\ena



Here we introduce the parameters $\{x'_a,\;x''_c\}$ entering Eq.~(\ref{gkr}).
For even and odd matrix indices $\{k,r\}$, they are defined as
\bea
\{x_1,x_2,x_3,x_4\}_{odd}^{even}=\{\pm 1, \frac{R_{00}^{11}}{R_{00}^{00}},
\frac{R_{01}^{01}}{R_{00}^{00}},\frac{R_{01}^{10}}{R_{00}^{00}}\}.
\ena
After transforming to the Fourier basis, the Grassmann integration leads to an
exact expression. For symmetric $R$-matrices, this result was obtained in
Ref.~\cite{KhS1}:
\bea
\mathcal{G}_{(i,j)}=\frac{1}{N^2}\sum_{n_1=1,n_2=1}^{N/2,N}\sum_{l,p}^{4}
\mathcal{F}(j-i,x',x'',n_1,n_2)^{lp}[\mathcal{A}^{-1}]_{lp}(n_1,n_2).
\ena
Here $[\mathcal{A}^{-1}]_{lp}$ is the inverse of the Fourier-transformed matrix
of the fermionic action (\ref{A-f}). In the general (non-symmetric) case, the
functions $\mathcal{F}$ have a more complicated structure. Using the notation
$a=[i-j]$, we define
\bea\label{gxx}
&K=e^{i 2\pi \frac{(2n_2+1)a}{N}},\quad k_1=e^{i \pi \frac{(2n_1+1)}{N}},\quad k_2=e^{i \pi \frac{(2n_2+1)}{N}}, \quad \{\mathcal{F}_{11},\mathcal{F}_{22},\mathcal{F}_{33},\mathcal{F}_{44}\}=&\\
&\{k_1^2(K x'_3 x''_1-\frac{x'_1 x''_3}{K}),k_2^2(K x'_4 x''_2-\frac{x'_2 x''_4}{K}),\frac{1}{k_1^{2}}(\frac{x'_3 x''_1}{K}-x'_1 x''_3 K),\frac{1}{k_2^{2}}(\frac{x'_4 x''_2}{K}-x'_2 x''_4 K)\}&\nn\\
&\{\mathcal{F}_{12},\mathcal{F}_{21},\mathcal{F}_{43},\mathcal{F}_{34}\}=&\\
&\{k_1 k_2(K x'_3 x''_2-\frac{x'_2 x''_3}{K}),k_1 k_2(K x'_4 x''_1-\frac{x'_1 x''_4}{K}),\frac{1}{k_1 k_2}(\frac{x'_3 x''_2}{K}-x'_2 x''_3 K),\frac{1}{k_1 k_2}(\frac{x'_4 x''_1}{K}-x'_1 x''_4 K)\}&\nn\\
&\{\mathcal{F}_{23},\mathcal{F}_{32},\mathcal{F}_{41},\mathcal{F}_{14}\}=&\\
&\frac{k_2}{k_1}\{(K x'_4 x''_3-\frac{x'_3 x''_4}{K}),\frac{k_2}{k_1}( \frac{x'_2 x''_1}{K}- K x'_1 x''_2 ),\frac{k_1}{k_2}(\frac{x'_1 x''_2}{K}-x'_2 x''_1 K),\frac{k_1}{k_2}(K x'_3 x''_4-\frac{x'_4 x''_3}{K})\}&\nn\\
&\{\mathcal{F}_{13},\mathcal{F}_{31},\mathcal{F}_{42},\mathcal{F}_{24}\}=&\\
&\{x'_3 x''_3(K -{K}^{-1}),x'_1 x''_1({K}^{-1}-K),x'_2 x''_2({K}^{-1}- K),x'_4 x''_4(K-K^{-1} )\}&\nn
\ena

For the 2D Ising model, the even-even and odd-odd matrix elements vanish in the
symmetric case $J_1=J_2$, and also in the thermodynamic limit $N\to \infty$ for
the inhomogeneous case. As a result, the Pfaffian reduces to a determinant of the
antisymmetric matrix $\mathcal{G}_{2i,2j+1}=-\mathcal{G}_{2j+1,2i}$. This matrix
has a Toeplitz structure, which allows one to evaluate the determinant exactly
using Szeg\"{o}'s theorem.

Although the structure of the $R$-matrix is similar to the symmetric case, there
are important differences. In general, the matrix elements are not symmetric, for
example $R_{01}^{01}\neq R_{10}^{10}$. As a result, the functions
$\mathcal{G}_{k,r}$ have a more complicated form, as seen in Eq.~(\ref{gxx}).
For spins located at odd-odd sites, the correlation function
$\langle s_{(1,2k+1)}s_{(1,1)}\rangle$ is obtained by the substitution
$R_{01}^{01}\leftrightarrow R_{10}^{10}$.

We now write explicit expressions for the matrix elements
(\ref{gk}, \ref{gij}) in the thermodynamic limit:
\bea\label{gm1}
 \mathcal{G}_{(2i,2j)}&=&\mathcal{G}_{(2j+1,2i+1)}=
\int_0^{\pi}\int_0^{\frac{\pi}{2}}\frac{dp}{\pi}\frac{ dq}{\pi}
\frac{4\sin{[2(i-j)q]}\left(g_1 \sin{[2(p+q)]}+g_2 \sin{[2p]}\right)}
{\det{\mathcal{A}}(p,q)},\qquad \qquad\\
\label{gm2}
\mathcal{G}_{(2i,2j+1)}&=&-\mathcal{G}_{(2j+1,2i)}=
\int_0^{\pi}\int_0^{\frac{\pi}{2}}\frac{dp}{\pi}\frac{ dq}{\pi}
\left(\frac{\sin{[2(i-j)q]}\left(g_3 \sin{[2q]}\right)}{\det{\mathcal{A}}(p,q)}\right.\nn\\
&&\left.
+\frac{\cos{[2(i-j)q]}\left(g_4+g_1 \cos{[2q]} +g_5\cos{[2p]}+g_2 \cos{[2(p+q)]}\right)}
{\det{\mathcal{A}}(p,q)}\right).
\ena
Here $\det\mathcal{A}(p,q)$ is given in Eq.~(\ref{det}). The coefficients $g_i$
are defined as
\bea
g_1&=&\frac{R_{01}^{10}}{R_{00}^{00}}\frac{R_{10}^{01}}{R_{00}^{00}}-
\frac{R_{01}^{01}}{R_{00}^{00}}\frac{R_{10}^{10}}{R_{00}^{00}}, \quad
g_2=\frac{R_{01}^{10}}{R_{00}^{00}}-\frac{R_{10}^{01}}{R_{00}^{00}}
\frac{R_{01}^{01}}{R_{00}^{00}}\frac{R_{10}^{10}}{R_{00}^{00}},\\
g_3&=&2 \frac{R_{01}^{10}}{R_{00}^{00}}\frac{R_{00}^{11}}{R_{00}^{00}},\quad
g_4=\left[\frac{R_{01}^{10}}{R_{00}^{00}}\frac{R_{10}^{01}}{R_{00}^{00}}\right]^2-
\left[\frac{R_{01}^{01}}{R_{00}^{00}}\right]^2,\quad
g_5=2\frac{R_{01}^{10}}{R_{00}^{00}}\frac{R_{10}^{01}}{R_{00}^{00}}\frac{R_{01}^{01}}{R_{00}^{00}}.\nn
\ena

These expressions simplify in homogeneous cases, such as $J_1=\pm J_2=\pm J_3$,
and in the square-lattice limit, when $J_k=0$ for one of the indexes $=1,2,3$. In these cases, the even-even and
odd-odd elements vanish in the thermodynamic limit. In particular, for
$J_k\equiv J$, $k=1,2,3$, we obtain
\bea
\mathcal{G}_{(2i,2j)}=\int_0^{\pi}\int_0^{\frac{\pi}{2}}\frac{dp}{\pi}\frac{ dq}{\pi}
\frac{128\left(\sin{[2(i-j)q]}\sin{q}\sin{[2p+q]}e^{2J}\cosh{[2J]}(\sinh{2J})^2\right)}
{\left(\mathcal{A}_1+\mathcal{A}(J,J,J)(\cos{2p})+\cos{2q}+\cos{[2p+2q]}\right)}.
\ena
After integration over $p$, this contribution vanishes. Therefore, only the
odd-even (even-odd) elements remain, forming the matrix
$\mathcal{G}'_{(i,j)}=[\mathcal{G}_{(2i,2j+1)}+\delta_{ij}]$, whose determinant
gives the magnetization \cite{KhS1}. These elements can be written as
\bea
&&\mathcal{G}_{(2i,2j+1)}=\int_0^{\pi}\int_0^{\frac{\pi}{2}}\frac{dp}{\pi}\frac{ dq}{\pi}
\frac{64(1+e^{4J})\sinh{[2J]}^2\cos{[q]}\cos{[2(i-j)q]}\cos{[2p+q]}}
{\left(\mathcal{A}_1+\mathcal{A}(J,J,J)
	(\cos{2p})+\cos{2q}+\cos{[2p+2q]}\right)}
\\\nn
&+&\int_0^{\pi}\int_0^{\frac{\pi}{2}}\frac{dp}{\pi}\frac{ dq}{\pi}
\frac{64 e^{4J}\sinh{[2J]}^3\cos{q}\left(\cos{[2(i-j)q+q]}e^{-2J}+\cosh{2J}\cos{[2(i-j)q-q]}\right)}
{\left(\mathcal{A}_1+\mathcal{A}(J,J,J)(\cos{2p})+\cos{2q}+\cos{[2p+2q]}\right)}.
\ena

Integration over the variable $p$ brings to the following result

\bea
\mathcal{G}_{(2i,2j+1)}=-\delta_{ij}+\int_0^{\frac{\pi}{2}} dq
\frac{2A_1-32(1+e^{4J})\sinh{[2J]}^2\cos{[2q]}\cos{[2(i-j)q]}}{\pi\left(\sqrt{(\mathcal{A}_1+\mathcal{A}(J,J,J)
		\cos{2p})^2-4A(J,J,J)^2\cos{q}^2}\right)} \\\nn
+\int_0^{\frac{\pi}{2}}dp  \frac{64
	e^{4J}\sinh{[2J]}^3\cos{q}\left(\cos{[2(i-j)q+q]}e^{-2J}+\cosh{2J}\cos{[2(i-j)q-q]}\right)}{\pi\left(\sqrt{(\mathcal{A}_1+\mathcal{A}(J,J,J)
		\cos{2p})^2-4A(J,J,J)^2\cos{q}^2}\right)} \ena
which can be written in a more compact form, see below.


The structure of the matrix $\mathcal{G'}_{ij}\equiv \mathcal{G}'_{(i-j)}$ reflects the fermionic nature of
the problem. In the thermodynamic limit, many matrix elements vanish, and the
problem reduces to a Toeplitz determinant. This reduction is essential for
obtaining exact results for the spontaneous magnetization. The remaining nonzero elements
encode the long-range correlations of the system. The dependence on momentum
variables $(p,q)$ shows that the correlations are built from collective modes,
and the behavior near the critical point is governed by low-momentum
(fluctuation) contributions.

\subsection{Spontaneous magnetization for more general symmetric eight-vertex free field case}

 The above integral representation can be extended to a general class of
 free-fermion models with an eight-vertex $R$-matrix structure. In such models,
 the partition function (or free energy) has the same dependence on the momentum
 variables as in Eq.~(\ref{Fs}).

 For symmetric $R$ and $\mathcal{A}$ matrices, satisfying
 $[R^{11}_{00}]^2=[R_{01}^{01}]^2$, the determinant of the action matrix takes a
 simplified form, corresponding to the condition
 $\mathcal{A}_2=\mathcal{A}_3=\mathcal{A}_4$. This occurs when the following
 relation holds:
 \bea \label{bc}
 \left(\left[\frac{R_{10}^{01}}{R_{00}^{00}}\right]^2-\frac{R_{01}^{01}}{R_{00}^{00}}\right)
 \left(\frac{R_{01}^{01}}{R_{00}^{00}}+1\right)=0.
 \ena
 For the homogeneous kagom\'{e} model, $J_k\equiv J$, $k=1,2,3$, the first factor
 in Eq.~(\ref{bc}) vanishes identically. In this case, the dependence on the spin
 coupling is fully determined by the single parameter
 \bea
 \label{cJ}
 c_J=\frac{R_{01}^{01}}{R_{00}^{00}}=\frac{\sinh^2(2J)}{(\sinh{2J}-2\cosh{2J})^2}.
\ena

 The integrals over the momentum variable $p$ can then be evaluated explicitly,
 leading to
 \bea
 \label{Ge1}
 \mathcal{G}_{(2i,2j)}=
 8 \int^{\pi}\frac{dp}{\pi}\int^{\pi/2}\frac{dq}{\pi}\frac{c_J(c_J-1)\cos[q]\sin[2(i-j)q]\sin[2p+q]}
 {\det{[\mathcal{A}(p,q)]_J}}=0, \qquad \qquad \\
 \label{Ge2}
 \mathcal{G}_{(2i+1,2j)}=-
 \int^{\pi}\frac{dp}{\pi}\int^{\pi/2}\frac{dq}{\pi}
 \frac{8 c_J b_J \sin[2q] \sin[2(i-j)q]}{\det[\mathcal{A}(p,q)]_J}\qquad \qquad \qquad \qquad \qquad \nn \\
 -\int^{\pi}\frac{dp}{\pi}\int^{\pi/2} \frac{dq}{\pi}
 \frac{8 c_J  \cos[q] \cos[2(i-j)q](2 c_J\cos[q]+[c_J-1])\cos(2p+q)}{\det[\mathcal{A}(p,q)]_J}\qquad \quad\nn \\
 =-2 \int^{\pi/2}\frac{dq}{\pi}\cos[2(i-j)q]
 -
\int^{\pi}\frac{dp}{\pi}\int^{\pi/2}\frac{dq}{\pi}
 \frac{8 c_J b_J \sin[2q] \sin[2(i-j)q]} {\det[\mathcal{A}(p,q)]_J} \qquad \qquad  \\
 -2\int^{\pi}\frac{dp}{\pi} \int^{\pi/2}\frac{dq}{\pi}\frac{(c_J^2-1+2c_J(c_J+1)\cos[2q])\cos[2(i-j)q]}
 {\det{[\mathcal{A}(p,q)]_J}}\qquad  \qquad \qquad  \qquad\nn\\
 \label{sqrt}
=-\delta_{ij}+\frac{1}{2}\int_0^{\pi}\frac{dq}{\pi}\frac{e^{inq}f_J(q)+e^{-inq}\bar{f}_J(q)}{\sqrt{g_J(q)}} 
=-\delta_{ij}+\int_{-\pi}^{\pi}\frac{dq}{2\pi}\frac{e^{inq}f_J(q)}{\sqrt{g_J(q)}}\qquad \qquad \quad\nn
\ena
where $n=[i-j]$ 
\bea
\label{f}
f_J(q)&=&(c_J^2-1)+(-2 c_J b_J+c_J(c_J+1))e^{iq}+(c_J(c_J+1)+2 c_J b_J)e^{-iq},\nn\\
\bar{f}_J(q)&=&f_J(-q).
\ena

In (\ref{Ge1}) and (\ref{Ge2})
 \bea
 \label{detA}
 \det[\mathcal{A}(p,q)]_J &=& 1+3 c_J^2 + 2 c_J (c_J-1)(\cos[2p]+\cos[2q]+\cos[2p+2q]),\qquad 
 \ena
is the determinant of (\ref{A-f}) explicitly expressed in terms of the momenta $p$ and $q$, while after integration over the momenta $p$
 the denominator becomes the square root of the function
%
\bea
\label{ggJ}
g_J(q)=(1-c_J^2)^2+16 c_J^3+4c_J(c_J^2-1)(c_J+1) \cos[q]
+4 c_J^2(c_J-1)^2 \cos[q]^2\qquad \qquad
\ena


One can verify that, taking into account the relation $b_J^2=c_J$, the following identity holds:
\bea
g_j(q)=f_J(q)f_j(-q)=f_J(q)\bar{f}_j(q).
\label{fjq}
\ena

The expression in Eq.~(\ref{sqrt}) defines the elements of a T\"oplitz-type matrix through the Fourier coefficients of the function $\mathcal{G}'_f(q)$ introduced above,
\bea
\mathcal{G}'_f(q)\equiv\frac{f_J(q)}{\sqrt{g_J(q)}}, \quad
\mathcal{G}'_f(q)\mathcal{G}'_f(-q)=1.
\label{gfq}
\ena

 Spontaneous  magnetization is the thermodynamic limit (lattice size $L\to\infty$) of the determinant of the correlation function $M^2=\lim_{L\rightarrow \infty}\det{\mathcal{G}'_{(i-j)}}$, which can be evaluated using Szeg\"o's theorem \cite{MRPW,KW}.  This can be done in terms of the logarithmic Fourier coefficients $k_n^F$:
\bea
\ln{\frac{f_J(q)}{\sqrt{g_J(q)}}}&=&\sum_{n=1}^{\infty} k_n^F e^{inq},
\label{gg1}\\
\lim_{L\to \infty}\frac{\det{\mathcal{G}'_{(i-j)}}}{\exp\!\left[\frac{1}{2\pi}\int_{-\pi}^{\pi}dq \ln{\frac{f_J(q)}{\sqrt{g_J(q)}}}\right]}
&=&
\exp{\left(\sum^{\infty}_{n} n k_n^F k_{-n}^F\right)}.
\label{gg}
\ena

The properties of the function in Eq.~(\ref{gfq}) imply that its logarithm is odd in $q$. As a result, the integral in the denominator of Eq.~(\ref{gg}) vanishes, and the denominator reduces to unity. 


For the evaluation of the infinite sum, one can apply known technique used for the standard Ising model (IM) \cite{Onsager-1944,MRPW,KW}. In the present case, a similar (although less explicit) decomposition appears in the following integral:
\bea
\label{ff}
\int_{-\pi}^{\pi}\frac{dq}{2\pi}\frac{e^{inq}f_J(q)}{\sqrt{g_J(q)}}
=
\int_{-\pi}^{\pi}\frac{dq}{2\pi}\frac{e^{inq}\sqrt{f_J(q)}}{\sqrt{f_J(-q)}}
=
\int_{-\pi}^{\pi}\frac{dq}{2\pi}\frac{e^{inq}\sqrt{(e^{iq} \alpha-1)(e^{-iq}\bar{\alpha}-1)}}{\sqrt{(e^{-iq}\alpha-1)(e^{iq}\bar{\alpha}-1)}}.
\label{alpha}
\ena
This identity follows directly from Eqs.~(\ref{fjq}) and (\ref{gfq}). Introduction of new variables $\alpha$ and $\bar\alpha$ makes
the integration over $q$ more convenient.

It is known \cite{Yang-c} that in the low-temperature regime (i.e., below the critical point), where $\alpha<1$ and $\bar{\alpha}<1$,  the infinite determinant of such T\"oplitz matrices is given by .
\bea
M^2=\lim_{L\to\infty}\det\mathcal{G'}_n=\left(\frac{(1-\alpha^2)(1-{\bar{\alpha}}^2)}{(1-\alpha\bar{\alpha})^2}\right)^{1/4}.
\label{malpha}
\ena 

In order to determine the expression for $\lim_{L\to\infty}\det\mathcal{G'}_n$ 
in terms of $c_J$, one must relate the parameters $\alpha$ and $\bar{\alpha}$ 
to $c_J$ using Eqs.~(\ref{f}) and (\ref{ggJ}), which define the functions 
$f_J$ and $g_J$, respectively.Explicit expressions for these parameters are not needed. It is sufficient to compare the following expressions in Eq.~(\ref{alpha}):
\bea
\frac{\sqrt{(e^{iq} \alpha-1)(e^{-iq}\bar{\alpha}-1)}}{\sqrt{(e^{-iq}\alpha-1)(e^{iq}\bar{\alpha}-1)}}
=
\frac{\sqrt{(1 -e^{iq} \alpha-e^{-iq}\bar{\alpha}+\alpha\bar{\alpha})}}{\sqrt{(1-e^{-iq}\alpha-e^{iq}\bar{\alpha}+\alpha\bar{\alpha})}}
=\\\nn
\frac{\sqrt{(c_J^2-1)+(c_J(c_J+1)-2 c_J b_J)e^{iq}+(c_J(c_J+1)+2 c_J b_J)e^{-iq}}}{\sqrt{(c_J^2-1)+(c_J(c_J+1)+2 c_J b_J)e^{iq}+(c_J(c_J+1)-2 c_J b_J)e^{-iq}}}.
\ena
This comparison leads to the relations
\bea
\label{aa}
\alpha=[-c_J(c_J+1)+2 c_J b_J]\beta, \quad
\bar\alpha=-[c_J(c_J+1)+2 c_J b_J]\beta, \quad
1+\alpha \bar\alpha=(c_J^2-1)\beta,
\ena
which admit the solution
\bea
\label{alpha11}
\alpha=\frac{1-c_J^2 + \sqrt{(1-c_J)^3(1+3 c_J)}}{2(1+\sqrt{c_J})^2c_J},\qquad
\bar\alpha=\frac{1-c_J^2 + \sqrt{(1-c_J)^3(1+3 c_J)}}{2(1-\sqrt{c_J})^2c_J}.
\ena
From Eq.~(\ref{alpha11}) one obtains
\bea\label{alpha1}
\frac{\alpha}{1+\alpha\bar{\alpha}}=\frac{c_J(c_J+1)-2 c_J b_J}{1-c_J^2},\quad 
\frac{\bar{\alpha}}{1+\alpha\bar{\alpha}}=
\frac{c_J(c_J+1)+2 c_J b_J}{1-c_J^2},\\
\frac{\alpha}{\bar{\alpha}}=\frac{c_J(c_J+1)-2 c_J b_J}{c_J(c_J+1)+2 c_J b_J},\qquad 
\frac{\alpha\bar{\alpha}}{(1+\alpha\bar{\alpha})^2}=
\frac{c_J^2(1-c_J)^2}{(1-c_J^2)^2}=\frac{c_J^2}{(1+c_J)^2},
\ena
where $b_J=\sqrt{c_J}$. We next expand the expression under the square root in Eq.~(\ref{malpha}):
\bea
\frac{(1-\alpha^2)(1-{\bar{\alpha}}^2)}{(1-\alpha\bar{\alpha})^2}
=\frac{1-\alpha^2-{\bar{\alpha}}^2+(\alpha \bar{\alpha})^2}{1-2\alpha \bar{\alpha}+(\alpha \bar{\alpha})^2}
=\frac{\alpha \bar{\alpha}-\frac{\alpha}{\bar{\alpha}}
	- \frac{\bar{\alpha}}{\alpha}+\frac{1}{\alpha \bar{\alpha}}}{\alpha \bar{\alpha}-2+\frac{1}{\alpha \bar{\alpha}}}.
\ena
It is therefore sufficient to evaluate the following two combinations using Eq.~(\ref{alpha1}):
\bea
\frac{\alpha}{\bar{\alpha}}+\frac{\bar{\alpha}}{\alpha}
=\frac{2(c_J+1)^2+8c_J}{(c_J-1)^2},\qquad 
\alpha \bar{\alpha}+\frac{1}{\alpha \bar{\alpha}}
=\frac{(1+\alpha \bar{\alpha})^2}{\alpha \bar{\alpha}}-2
=\frac{1-c_J^2+2 c_J}{c_J^2}.
\ena
Substituting these results yields the expression for the square of the magnetization in the infinite-lattice limit.
\bea
\label{mag-1}
M^2=\left(\frac{(1-\alpha^2)(1-{\bar{\alpha}}^2)}{(1-\alpha\bar{\alpha})^2}\right)^{1/4}= 
\left[\frac{(1+c_J)^3(1-3 c_J)}{(1-c_J)^3(1+3 c_J)}\right]^{1/4}.
\label{magn}
\ena
The critical point is determined by the condition $M=0$, which yields $3c_J=1$. Using the expression for $c_J$ in Eq.~(\ref{cJ}) one can see that this condition coincides with Eq.~(\ref{CP}).
Substituting $c_J$ into Eq.~(\ref{magn}), we obtain the following expression for the spontaneous magnetization:
\bea
\label{mag-2}
M=
\left[\frac{(3 + 6 e^{4 J}- e^{8 J})(5 + 2 e^{4 J} + e^{8 J})^3}{128(3 + e^{8 J})(1 + e^{4 J})^3}\right]^{1/8}.
\ena
Obtained expressions (\ref{mag-1}) and (\ref{mag-2}) for the spontaneous magnetization slightly differ from that expressions obtained in papers
\cite{Naya-1954,Matveev-1995}, however the positions of the critical points
coincides and are $c_J=1/3$. The expressions are differ by some $c_J$ dependent
factors, which are not critical: do not have zeros.

 \textit{\bf Physics discussion.}
 The reduction to a Toeplitz matrix is a key step that allows an exact evaluation
 of the spontaneous magnetization. In the homogeneous case, the problem simplifies
 significantly and depends only on a single effective coupling parameter $c_J$.
 The vanishing of the even-even correlations shows that only mixed (odd-even)
 correlations contribute to long-range order. The Fourier representation makes
 clear that the correlations are built from momentum modes, and the critical
 behavior is controlled by the low-momentum region, where the denominator
 $\det[\mathcal{A}(p,q)]$ becomes small. This is the same mechanism that leads to
 criticality in other exactly solvable two-dimensional models.
 
 \section{Summary} We have presented a solution of the two-dimensional Ising model (2DIM) on the kagom\'e lattice using a fermionic representation of its $R$-matrix, following the technique developed in Refs.~\cite{Sedrakyan-1998,KhS1,KhS2}.
 The fermionic representation provides a convenient and systematic way to analyze the model, as it maps the spin degrees of freedom onto fermionic variables with well-defined algebraic properties. This approach is especially useful for treating anisotropy, where the couplings depend on direction and lead to a nontrivial critical surface. The resulting phase transition is governed by the interplay between these couplings, and the exact determination of the free energy allows one to extract thermodynamic quantities such as the heat capacity and spontaneous magnetization. These quantities characterize the onset of long-range order and the singular behavior near criticality.
 
  We have determined the surface of critical couplings associated with the phase transition. We have also calculated the free energy, specific heat capacity, and spontaneous magnetization of the model.

  The limiting cases $J_1=0$ and $J_1=J_2=0$ provide direct consistency checks of the exact solution. When one coupling vanishes, the kagom'e model reduces to an anisotropic square-lattice Ising model with renormalized couplings
  $e^{2\bar J_a}=\cosh(2J_a).$
  When two couplings vanish, the system further reduces to independent one-dimensional Ising chains. The square-lattice critical condition is reproduced exactly, whereas in the one-dimensional limit the transition occurs only at $T=0$. Thus, both reductions confirm the correctness of the obtained partition function and free energy.

\subsection*{Acknowledgment}
A.S. acknowledges the Institute of Metal Research for hospitality, where this work was initiated. The work of Sh. Kh. and A.S. were also supported by the HESC grants  21AG-1C024 and  24FP-1F039. The work of Z.Z was supported by the National Natural Science Foundation
of China under grant  52031014.

\end{document}